\documentclass[twocolumn,prb,color,psfig,superscriptaddress]{revtex4}
\usepackage{graphicx}
\usepackage{epsfig}
\begin{document}
\title{
Dynamical charge inhomogeneity and crystal-field fluctuations for 4f ions in
high-T$_c$ cuprates}
\author{A.S. Moskvin}
\author{N.V. Mel'nikova}
\affiliation{Ural State University, 620083, Ekaterinburg, Russia}
\date{\today}
\begin{abstract}
The main relaxation mechanism of crystal-field excitations in rare-earth ions
in cuprates is believed to be provided by the fluctuations of crystalline
electric field induced by a dynamic charge inhomogeneity generic for the doped
cuprates.
We address the generalized granular model as one of the model scenario for such
an ingomogeneity where the cuprate charge subsystem  remind that of Wigner
crystal with the melting transition and phonon-like positional excitation
modes. Formal description of R-ion relaxation coincides with that of recently
suggested magnetoelastic mechanism.

\end{abstract}
\maketitle

Inelastic neutron scattering (INS) spectroscopy is a powerful tool to determine
unambiguously the Stark multiplet structure and crystal-field (CF) potential in
rare-earth (R) based high-T$_c$ superconducting materials such as
Y$_{1-x}$R$_x$Ba$_2$Cu$_3$O$_{6+y}$.\cite{Mesot-JS-97,Mesot-NS-98} This
technique provides detailed information on the electronic ground state of the
R-ions which is important to understand the thermodynamic magnetic properties
as well as the observed coexistence between superconductivity and long-range
magnetic ordering of the R-ion sublattice at low temperatures. Moreover, INS
spectroscopy  is addressed to be a powerful tool for a quantitative monitoring
the decay of the antiferromagnetic state of the parent compound as well as the
evolution of the superconducting state upon doping, since the linewidths of CF
transitions are believed to directly probe the  electronic susceptibility.
The relaxation behavior appears to be extremely dependent upon the energy at
which the  susceptibility is being probed.
The crystal-field INS spectroscopy is widely used to reveal the opening of an
electronic gap in the normal state of underdoped superconductors
\cite{Mesot-JS-97} and examine its anisotropy.
\cite{Mesot-EPL-98,Boothroyd-PRL-96} Recently, the Ho${}^{3+}$ CF-INS
spectroscopy was used to investigate the oxygen and copper isotope effects on
the pseudogap in the high-temperature superconductors Ho-124 and
(LaHoSr)$_2$CuO$_4$.\cite{Rubio-PRL-00,Rubio-PRB-02}
However, the mechanism of the relaxation of rare-earth ions in cuprates becomes
the issue of hot debates,\cite{Boothroyd-PRB-01,Lovesey-PRB-01} that questions
the current interpretation of information detected by INS spectroscopy.

In the normal state the excited crystal-field
levels of R-ion interact with phonons,  spin fluctuations, and charge
carriers. These interactions limit the life-time
of the excitation; thus the observed crystal-field transitions
exhibit line broadening. Similarly the conventional Fermi-liquid metals the
interaction with the charge carriers is by far considered to be  the dominating
relaxation mechanism in cuprates. This interaction is usually assumed to be
isotropic exchange coupling with effective spin-Hamiltonian
$H_{ex}=-2I(g_{J}-1)({\bf s}\cdot {\bf J})$, where $I$ is exchange integral
that should be nearly independent of the particular R-ion under consideration,
$g_{J}$ Lande factor, ${\bf s}$ is the spin momentum of a charge carrier, and
${\bf J}$  the total momentum of  R-ion. Such a scenario seems to be a rather
natural, if taking into account the predominant spin channel of neutron
scattering.  The detailed theory of the respective relaxation mechanism was
developed by Becker, Fulde and Keller (BFK-model).\cite{BFK}  The corresponding
intrinsic linewidth appears to increase almost linearly
with temperature ($\Gamma (T)\propto \rho ^{2}T$) according to the well-known
Korringa law.\cite{Korringa} Here $\rho$ is the {\it coupling constant}: $\rho
= I(g_{J}-1)N(E_{F})$, where $N(E_{F})$ is the density of states (DOS) at the
Fermi level. Namely the deviation from a linear temperature dependence at low
temperatures has been usually interpreted in terms of the opening of a
(pseudo)gap and the associated reduction in the damping. Fitting the
high-temperature linewidth data in frames of simple or modified Korringa law
one obtains the values of {\it coupling constant} which typically   vary from
0.003 to 0.006.
\cite{Mesot-JS-97,Mesot-EPL-98,Boothroyd-PRL-96,Rubio-PRL-00,Rubio-PRB-02}

It should be emphasized that the spin channel of relaxation implies directly
the relevance of the Fermi liquid scenario for cuprates ignoring many
signatures of non-Fermi liquid behavior. However, it is unlikely to be a
shortcoming of either model approach whether it were  intrinsically
self-consistent. However, the spin-exchange model has a number of visible
inconsistencies, firstly as concerns the magnitude of {\it coupling constant}.
Indeed, a linear temperature dependence of the relaxation time above T$_c$
observed in EPR studies of S-ion Gd$^{3+}$ in YBa$_2$Cu$_3$O$_7$  after
Korringa fitting yields the
  magnitude of exchange integral $I\approx 3\times 10^{-4}$ eV
(Ref.\onlinecite{Shaltiel}) that directly points to an unrealistically big
values of spin coupling constants $\rho$ found in all the INS experiments on CF
transitions. Some problems exist with a Lande factor $\propto (g_{J}-1)$
scaling. So, Mukherjee {\it et al}.\cite{1} when studying the system
Y$_{1-x}$R$_{x}$Ba$_{2}$Cu$_{3}$O$_{6+y}$ (R=Er, Ho, Tm) found $ |\frac {\rho
(Tm)}{\rho (Ho)}| \simeq 2$,
 instead of theoretically expected $\frac{(g_{Tm}-1)}{(g_{Ho}-1)}
=\frac{2}{3}$, and
 $|\frac {\rho (Tm)}{\rho (Er)}| \simeq 4.5$,
 instead of expected $\frac{(g_{Tm}-1)}{(g_{Er}-1)} =\frac{5}{6}$.
This clear disagreement evidences against exchange mechanism.
 The spin-exchange scenario fails to explain the "strange" doping dependence of
Tm$^{3+}$ relaxation in Tm-123 \cite{Osborn} and   Nd$^{3+}$
relaxation in (LaSrNd)$_2$CuO$_4$.\cite{Roepke}

Finally, Staub {\it et. al.}\cite{Staub} have found that the
Lorentzian linewidth of the quasi-elastic neutron scattering for
Tb$^{3+}$ in YBa$_2$Cu$_3$O$_7$  can be properly described by a
simple $(exp(\Delta /k_{B}T)-1)^{-1}$ law typical for Orbach
processes governed by lattice vibrations. They have  shown that such
an interpretation also describes the results obtained earlier on
Ho$^{3+}$ and Tm$^{3+}$. They conclude that the interactions with
the charge carriers are negligible and that the interactions with
the lattice vibrations are responsible for the relaxation behaviour
of the 4f electrons in cuprates. Therefore, the INS results which
claim to probe the superconducting gap or the pseudo-gap should be
re-examined in terms of Orbach processes. Similar conclusion was
drawn by Roepke {\it et. al.}\cite{Roepke} for Nd$^{3+}$ relaxation
in (LaSrNd)$_2$CuO$_4$. Lovesey and Staub \cite{Lovesey} have shown
that the dynamic properties of the lanthanide ions (Tb$^{3+}$,
Ho$^{3+}$, and Tm$^{3+}$ are adequately described by a simple
three-state model not unlike the one introduced by Orbach for the
interpretation of electron paramagnetic resonance signals from a
lanthanide ion in dilute concentration in a salt.
 The cross section for inelastic scattering of neutrons by the lanthanide ion
is derived by constructing a pseudospin S=1 model  and treating the
magnetoelastic interaction as a perturbation on the three crystal-field states.
 The scattering of neutrons is thus a quasielastic process and the relaxation
rate is proportional to $(exp(\Delta /k_{B}T)-1)^{-1}$, where $\Delta $  is the
energy of the  intermediate crystal-field state at which the density of phonon
states probed.
However, this very attractive scenario also faces some visible difficulties
with the explanation, for instance,  of the unusual nonmonotonic temperature
dependences and  too large oxygen isotope effect in the INS spectra of Ho-124
and Ho-214 systems,\cite{Rubio-PRL-00,Rubio-PRB-02} some doping dependences in
Nd-214 system.\cite{Roepke}

On comparing two mechanisms we should underline their difference which seems to
be  of primary importance: the spin-channel mechanism takes into account the
fluctuations of {\it effective magnetic field} on R-ion, while the phonon
(magnetoelastic) mechanism deals with that of {\it electric field}.
Moreover, the conventional spin-channel mechanism actually  probes {\it spin
fluctuations} rather than {\it charge fluctuations}, albeit its contribution to
the linewidth $\Gamma (T)\propto (I\,N(E_{F})^2$ is believed to strongly depend
on the density of carriers. However, this relationship is derived in frames of
Fermi liquid scenario, and would be modified, if one addresses the typical
antiferromagnetic insulating state.
Interestingly, that in Refs.\onlinecite{Staub,Lovesey,Lovesey-PRB-01} the
phonon (magnetoelastic) mechanism is addressed as an alternative to the charge
fluctuations. As an example, the authors point to insulating materials where
"...the density of carriers   is essentially zero...",\cite{Lovesey-PRB-01}
that forbids the charge  fluctuation channel of relaxation.

We would like to emphasize the fact that both groups of researchers
underestimate the role of the R-ion relaxation due to  a conventional {\it
spinless charge  fluctuation channel}. Indeed, the CF Hamiltonian for R-ion in
cuprate can be written in its standard form as follows:
$$
H_{CF}=\sum_{k=2,4,6}\sum_{-k\leq q\leq k}B^{*}_{kq}{\hat O}^{q}_{k}.
$$
 Here ${\hat O}^{q}_{k}$ are Stevens equivalent operators,
$B_{kq}=b_{kq}\langle r^{k}\rangle \gamma _{k}$, where $b_{kq}$ are CF
parameters, $\gamma _{2}=\alpha$,$\gamma _{4}=\beta$, $\gamma _{6}=\gamma$
($\alpha ,\beta ,\gamma$ are Stevens parameters);
$$
b_{kq} =\langle b_{kq}\rangle +\Delta b_{kq},
$$
which may be expressed  in frames of the well-known point-charge model as
follows:
$$
\Delta b_{kq}=\sum_{i}\frac{qC^{k}_{q}({\bf R}_{i})}{R_{i}^{k+1}}({\hat
n}_{i}(t)-\langle n_{i}\rangle),
$$
where $C^{k}_{q}$ is the spherical harmonics, ${\hat n}_{i}(t)$ the charge
number operator. Conventional metals are characterized by a very short-time
charge dynamics that allows to neglect the contribution of charge fluctuations
to the relaxation  of R-ions in the low-energy range of CF energies, and deal
with a mean homogeneous charge distribution.
An altogether different picture emerges in the case of cuprates where we deal
with various manifestations of static and dynamic charge inhomogeneity(see,
e.g. Refs.\onlinecite{Dagotto,Emery} and references herein). Moreover, the INS
spectroscopy of CF excitations itself yields the impressive picture of charge
inhomogeneity in 123 system,\cite{Mesot-JS-97,Mesot-NS-98} where it
  was found that the observed CF spectra separate into different local
components whose spectral weights distinctly depend on the doping level, i.e.,
there is clear experimental evidence for cluster formation. The onset of
superconductivity can be shown to result from percolation which means that the
superconductivity is an inhomogeneous materials property.
 It seems probable that the dynamical rearrangement of the charge system at the
temperatures above T$_c$ somehow  affects  the R-ion relaxation.

At present the stripe model of inhomogeneity \cite{Emery} became the most
popular. This model is based on the more universal idea of  topological phase
separation, when the doped particles are assumed to localize inside the domain
walls of a bare phase.

Below we address one of the topological phase separation scenario which may be
termed as {\it a generalized granular model for doped cuprates}. We assume that
the  CuO$_2$ layers in parent cuprates may gradually loose its stability under
electron/hole doping, while a novel self-organized multigranular 2D phase  is
believed to become stable.
Such a situation resembles in part that of granular superconductivity.

 The most probable possibility is that every micrograin accumulates one or two
particles. Then, the number of such entities in a multigranular texture
nucleated with doping  has to  nearly linearly depend  on the doping.
   Generally speaking, each individual
micrograin may be characterized by its position,  nanoscale size, and the
orientation of U(1) degree of freedom.
In contrast with the uniform states the phase of
the superfluid order parameter for micrograin is assumed to be unordered.
The granular structure must be considered as being largely dynamic in
nature.

In the long-wavelength limit the off-diagonal ordering can be described by an
effective Hamiltonian in terms of  U(1) (phase) degree of freedom associated
with each micrograin. Such a Hamiltonian
 contains a repulsive, long-range Coulomb part and a
short-range contribution related to the phase degree of freedom. The
latter term can be written out in the standard for the $XY$ model form of a
so-called Josephson coupling
\begin{equation}
H_J = -\sum_{\langle i,j\rangle}J_{ij}\cos(\varphi _{i}-\varphi _{j}),
\end{equation}
where $\varphi _{i},\varphi _{j}$ are global phases for micrograins centered at
points $i,j$, respectively, $J_{ij}$ Josephson coupling parameter. Namely the
Josephson coupling gives rise to the long-range ordering of the phase of the
superfluid order parameter in such a multi-center texture. Such a Hamiltonian
represents a starting point for the analysis of disordered superconductors,
granular superconductivity, insulator-superconductor transition with $\langle
i,j\rangle$ array of superconducting islands with phases $\varphi _{i},\varphi
_{j}$.

To account for Coulomb interaction and allow for quantum corrections we should
introduce into effective Hamiltonian  the charging energy \cite{Kivelson}
$$
H_{ch}=-\frac{1}{2}q^2 \sum_{i,j}n_{i}(C^{-1})_{ij}n_{j}\, ,
$$
where $n_{i}$ is a  number operator for particles bound in $i$-th micrograin;
it
is quantum-mechanically conjugated to $\varphi$: $n_{i}=-i \partial /\partial
\varphi
_{i}$, $(C^{-1})_{ij}$ stands for  the capacitance matrix, $q$ for a particle
charge.

 Such a system appears to reveal a tremendously rich quantum-critical
structure. In the absence of disorder, the
$T=0$ phase diagram of the multigranular system implies either triangular, or
square crystalline arrangements
 with possible melting transition to a  liquid.
 It should be noted that analogy with charged $2D$ Coulomb gas
implies the Wigner crystallization of  multigranular system with Wigner
crystal (WC) to Wigner liquid melting transition, respectively. Naturally, that
the
additional degrees of freedom for micrograin provide a richer physics of
such lattices. For a system  to be an insulator, disorder is required, which
pins the multigranular system
and also causes the crystalline order to have a finite correlation length.
Traditional approach to a Wigner crystallization implies the formation of a WC
for densities lower than a critical density, when the Coulomb energy dominates
over the kinetic energy. The effect of quantum fluctuations leads to a
(quantum) melting of the solid at high densities, or at a critical lattice
spacing. The critical properties of a two-dimensional lattice without any
internal degree of freedom are successfully described  applying the BKT theory
to dislocations and disclinations of the lattice, and proceeds in two steps.
The
first implies the transition to a liquid-crystal phase with short-range
translational order, the second does the transition to isotropic liquid.
 For such a system provided the micrograin
positions  fixed at all temperatures, the long-wave-length physics would be
described by an antiferromagnetic $XY$ model with expectable BKT transition and
gapless $XY$ spin-wave mode.

The low temperature physics in a multigranular system is  governed by an
interplay of two BKT transitions,  for the U(1) phase  and positional degrees
of
freedom, respectively. \cite{Timm}
Dislocations  lead to a mismatch in the U(1)
degree of freedom, which makes the dislocations bind fractional vortices and
lead to a coupling of translational and phase excitations. Both BKT
temperatures
either coincide (square lattice) or the melting one is higher (triangular
lattice).\cite{Timm}

 Quantum fluctuations can substantially affect these
results. Quantum melting can destroy U(1) order at sufficiently
low densities where the Josephson coupling becomes exponentially small. Similar
situation is expected to take place in the vicinity of
structural transitions in a multigranular crystal. With increasing the
micrograin density
the quantum effects  result in a significant lowering of the melting
temperature as compared with classical square-root dependence.
The resulting melting temperature can reveal  an oscilating behavior as a
function of particle density with zeros at the critical (magic) densities
associated with structural phase transitions.

In terms of our model, the positional order corresponds to an incommensurate
charge density wave, while the U(1) order does to a superconductivity. In other
words, we arrive at a subtle interplay between two orders. The superconducting
state evolves from a charge order with $T_C \leq T_m$, where $T_m$ is the
temperature of a melting transition which could be termed as a temperature of
the opening of the insulating gap (pseudo-gap!?).

The normal modes of a dilute  multigranular system
include the pseudo-spin waves
propagating in-between the micrograins, the positional fluctuations, or
quasi-phonon
modes,  which are gapless in  a pure system, but gapped  when
the lattice is pinned, and, finally,  fluctuations in the U(1) order parameter.

    The orientational
   fluctuations of the multigranular system are governed by the gapless
   $XY$ model.\cite{Green} The relevant model description is most familiar as
an effective   theory of the Josephson junction array. An important feature of
the model is that it displays a quantum-critical point.

The low-energy collective excitations of multigranular
liquid includes an usual longitudinal acoustic phonon-like branch.
The liquid crystal phases differ from the isotropic liquid in that they have
massive topological excitations, {\it i.e.}, the disclinations.
One should note that the liquids do not support transverse modes, these could
survive in a liquid state only as overdamped modes.  So that it is reasonable
to assume that solidification of the skyrmion lattice would be accompanied by a
stabilization of transverse phonon-like modes with its sharpening below melting
transition.
In other words an instability of transverse phonon-like modes signals the
onset of melting.
The phonon-like modes in skyrmion crystal have much in common with usual phonon
modes, however, due to electronic nature these can hardly  be detected if any
by inelastic neutron scattering.

A generic property of the positionally ordered skyrmion configuration is the
sliding mode which is usually pinned by the disorder. The depinning of sliding
mode(s) can be detected in a low-frequency and low-temperature optical
response.

It should be noted that as regards the CF fluctuations, there is no principal
difference between the contributions of real phonon modes and quasi-phonon
modes of a multigranular system. Moreover,  it is worth noting that the charge
inhomogeneity in a multigranular  system is proned to be closely coupled with
lattice structural  distorsions. However, the stabilization of transverse
phonon-like modes in multigranular  system that accompanies its solidification
at the temperatures above T$_c$ may strongly affect the CF relaxation due to a
mechanism  identical to that of proposed by Lovesey and Staub (magnetoelastic
mechanism). In a sense, such a conclusion reconciles   "old" spin-fluctuation
\cite{Mesot-JS-97,Mesot-NS-98} and "new" phonon \cite{Staub,Lovesey} approaches
to the INS spectroscopy of cuprates with R-ions.

  Concluding we argue  that the
crystal-field fluctuations induced by the dynamic charge inhomogeneity  in
copper-oxygen subsystem may be one of the main origin of the broadening of the
linewidth of  CF transitions for 4f ions in high-T$_c$ cuprates providing the
detection of a charge rearrangement accompanying the approach to  T$_c$.

 We acknowledge the support by  SMWK Grant, INTAS Grant No. 01-0654, CRDF Grant
No. REC-005, RME Grant No. E 02-3.4-392 and No. UR.01.01.042, RFBR Grant No.
01-02-96404. A.S.M. has benefited from stimulating discussions with A.T.
Boothroyd and J. Mesot.

\end{document}